\documentclass[pra,aps,twocolumn,10pt,showpacs,groupedaddress,superscriptaddress,floatfix,notitlepage]{revtex4-2}
\usepackage{amsmath,amsfonts,amssymb,graphics,graphicx,epsfig,color,times, braket}
\usepackage[utf8x]{inputenc}
\usepackage{color}
\usepackage{bbm, dsfont}
\usepackage{float}
\floatstyle{plaintop}
\restylefloat{table}
\usepackage{subfigure}
\usepackage{hyperref}
\usepackage{mathrsfs}
\usepackage{verbatim}
\usepackage{physics}
\usepackage{booktabs}
\usepackage{float}
\usepackage{silence}
\begin{document}
\title{Steady-state phases and interaction-induced depletion in a driven-dissipative chirally-coupled dissimilar atomic array}

\author{Shao-Hung Chung}
\thanks{These two authors contributed equally}
\affiliation{Department of Physics and Center for Quantum Science and Engineering,
National Taiwan University, Taipei 10617, Taiwan}
\affiliation{Institute of Atomic and Molecular Sciences, Academia Sinica, Taipei 10617, Taiwan}
\author{I Gusti Ngurah Yudi Handayana}
\thanks{These two authors contributed equally}
\affiliation{Molecular Science and Technology Program, Taiwan International Graduate Program, Academia Sinica, Taiwan}
\affiliation{Department of Physics, National Central University, Taoyuan City 320317, Taiwan}
\affiliation{Institute of Atomic and Molecular Sciences, Academia Sinica, Taipei 10617, Taiwan}
\author{Yi-Lin Tsao}
\affiliation{Institute of Atomic and Molecular Sciences, Academia Sinica, Taipei 10617, Taiwan}
\author{Chun-Chi Wu}
\affiliation{Institute of Atomic and Molecular Sciences, Academia Sinica, Taipei 10617, Taiwan}
\author{G.-D. Lin}
\affiliation{Department of Physics and Center for Quantum Science and Engineering,
National Taiwan University, Taipei 10617, Taiwan}
\affiliation{Physics Division, National Center for Theoretical Sciences, Taipei 10617, Taiwan}
\affiliation{Trapped-Ion Quantum Computing Laboratory, Hon Hai Research Institute, Taipei 11492, Taiwan}
\author{H. H. Jen}%
\email{sappyjen@gmail.com}
\affiliation{Institute of Atomic and Molecular Sciences, Academia Sinica, Taipei 10617, Taiwan}
\affiliation{Molecular Science and Technology Program, Taiwan International Graduate Program, Academia Sinica, Taiwan}
\affiliation{Physics Division, National Center for Theoretical Sciences, Taipei 10617, Taiwan}

\date{\today}
\renewcommand{\r}{\mathbf{r}}
\newcommand{\f}{\mathbf{f}}
\renewcommand{\k}{\mathbf{k}}
\def\p{\mathbf{p}}
\def\q{\mathbf{q}}
\def\bea{\begin{eqnarray}}
\def\eea{\end{eqnarray}}
\def\ba{\begin{array}}
\def\ea{\end{array}}
\def\bdm{\begin{displaymath}}
\def\edm{\end{displaymath}}
\def\red{\color{red}}
\pacs{}
\begin{abstract}
A nanophotonic waveguide coupled with an atomic array forms one of the strongly-coupled quantum interfaces to showcase many fascinating collective features of quantum dynamics. In particular for a dissimilar array of two different interparticle spacings with competing photon-mediated dipole-dipole interactions and directionality of couplings, we study the steady-state phases of atomic excitations under a weakly-driven condition of laser field. We identify a partial set of steady-state phases of the driven system composed of combinations of steady-state solutions in a homogeneous array. We also reveal an intricate role of the atom at the interface of the dissimilar array in determining the steady-state phases and find an alteration in the dichotomy of the phases strongly associated with steady-state distributions with crystalline orders. We further investigate in detail the interaction-induced depletion in half of the dissimilar array. This blockaded region results from two contrasting interparticle spacings near the reciprocal coupling regime, which is evidenced from the analytical solutions under the reciprocal coupling. Our results can provide insights in the driven-dissipative quantum phases of atomic excitations with nonreciprocal couplings and pave the avenues toward quantum simulations of exotic many-body states essential for quantum information applications. 
\end{abstract}
\maketitle
\section{Introduction}

A driven-dissipative quantum system \cite{Diehl2008, Kraus2008, Verstraete2009, Baumann2010, Weimer2010, Diehl2010, Barreiro2011} provides unprecedented opportunities to explore nonequilibrium phase transitions \cite{Diehl2010} and to create strongly-correlated steady states useful for quantum information processing \cite{Verstraete2009}. This relies on an interplay or a competition between dissipations and interaction strengths, from which novel quantum many-body states and associated rich dynamical phenomena can emerge. Recently, an intriguing atom-waveguide QED platform \cite{Lodahl2017, Chang2018, Samutpraphoot2020, Masson2020, Jen2020_phase, Fayard2021, Kim2021, Sheremet2023}, a distinct class of open quantum systems \cite{Vetsch2010, Thompson2013, Goban2015, Corzo2019, Kim2019, Jen2020_disorder, Dordevic2021, Iversen2022, Fedorovich2022}, has showcased nontrivial collective radiations \cite{Henriet2019, Zhang2019, Ke2019, Albrecht2019, Needham2019, Jen2020_subradiance, Mahmoodian2020, Jen2021_bound, Pennetta2022, Pennetta2022_2} and long-range quantum correlations \cite{Tudela2013, Mahmoodian2018, Downing2020, Jeannic2021, Jen2022_correlation} owing to the strong couplings between atoms and the guided modes \cite{Douglas2015, Solano2017}, and the emergent nonreciprocity \cite{Mitsch2014, Ramos2014, Pichler2015, Lodahl2017} in the bidirectional couplings controlled via external magnetic fields. 

This controlled and effective nonreciprocal coupling between atoms has been implemented in artificial quantum emitters \cite{Luxmoore2013, Arcari2014, Yalla2014, Sollner2015, Roushan2017, Wang2019}, atom-nanofiber systems \cite{Mitsch2014, Solano2017, Corzo2019, Sayrin2015}, and diamond nanophotonics platforms \cite{Sipahigil2016, Bhaskar2017}. Within these strongly-coupled systems, it is the photon-mediated and long-range dipole-dipole interaction \cite{Solano2017} that leads to significant quantum correlations and allows many intriguing applications in routing or interfering photons, useful for integrated quantum network and scalable quantum computation \cite{Sollner2015}. An extra degree of freedom in manipulating the directionality of couplings \cite{Mitsch2014} thus offers new angles in studying quantum dynamics in these quantum interfaces and provides novel applications in quantum simulations or quantum computation in the next-generation nanophotonic devices. 

Recent efforts have focused on integrating dissimilar arrays with clean and disordered zones \cite{Leonard2023} or with disparate interparticle spacings \cite{Wu2023, Yudi2023}. The former intends to reveal the mechanism of many-body delocalization in the Bose-Hubbard model, while the latter investigates the Anderson-like localization to delocalization transition and atomic excitation trapping effect. This dissimilar array is less explored especially under the driven-dissipative condition, where distinct steady-state phases can emerge with an interplay between the directionality of couplings and the strength of photon-mediated dipole-dipole interaction. Here we study the steady-state phases in a driven-dissipative dissimilar atomic array with chiral couplings, where a plethora of steady-state phases are uncovered with features resembling some of the steady-state phases in a homogeneous array. A competition between interparticle spacings and directionality of couplings further results in an interaction-induced depletion of atomic excitations, in huge contrast to the homogeneous array under a uniform and weak laser excitation. We investigate this unique steady-state phase with exploration of various parameters and locate the parameter region that supports the interaction-induced depletion feature, along with explanations from analytical calculations in the reciprocal coupling regime. Our findings can offer valuable insights into the driven-dissipative and interaction-induced quantum phases of atomic excitations with nonreciprocal couplings and open new avenues for simulations of exotic quantum states useful for quantum information applications. 

The paper is organized as follows. In Sec. II, we introduce the theoretical model of driven-dissipative dissimilar atomic array with two disparate interparticle spacings and nonreciprocal couplings. In Sec. III, we present the steady-state phases of atomic excitations. We further investigate the interaction-induced depletion in the dissimilar array in Sec. IV, where we identify the parameter regions that manifest the significant excitation depletion. Finally, we discuss and conclude in Sec. V.

\begin{figure}[t]
\centering
\includegraphics[width=0.48\textwidth]{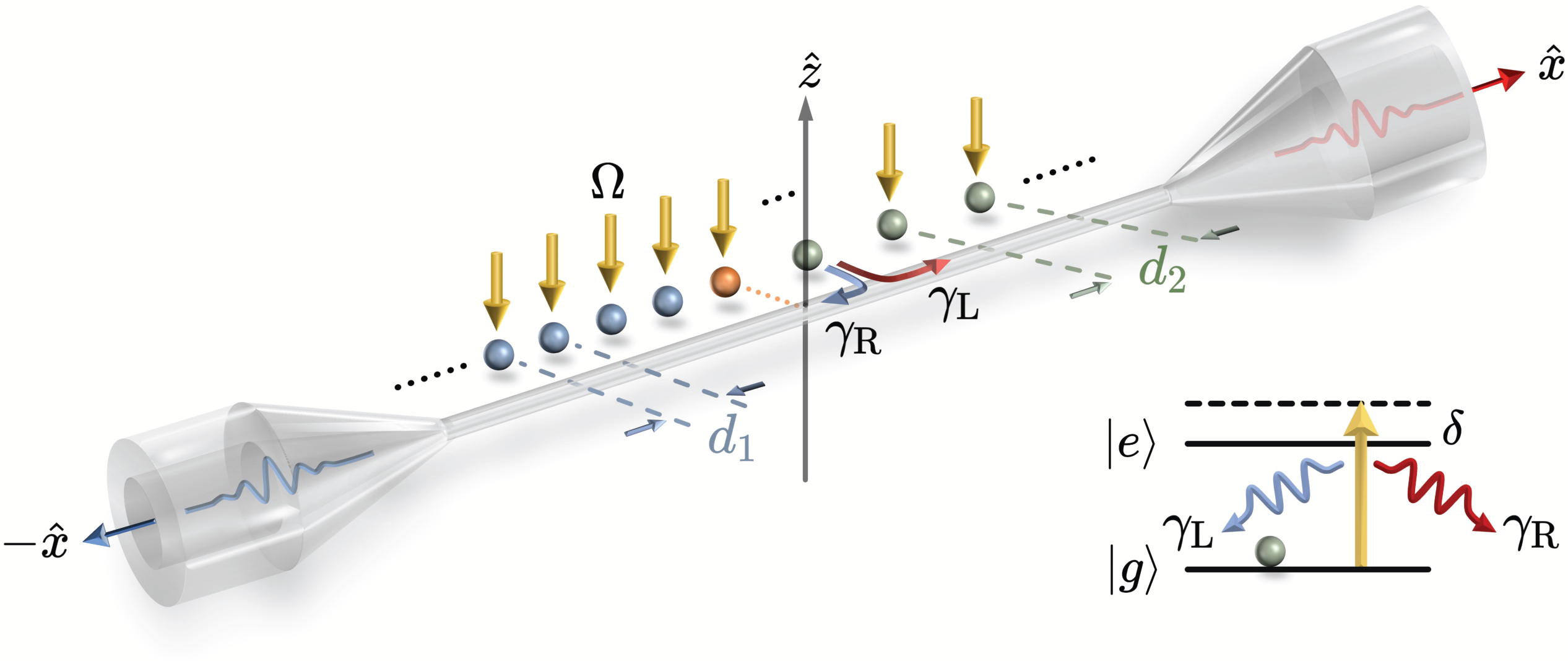}
\caption{Schematic illustration of weakly-driven dissimilar chirally-coupled atomic array. The system comprises two-level quantum emitters with effective nonreciprocal decay rates, $\gamma_{\text{L}} \neq \gamma_{\text{R}}$, which can be facilitated by the guided modes on a waveguide with external magnetic fields, and is driven weakly by a uniform laser field with a Rabi frequency $\Omega$ and a detuning $\delta$ from a side excitation. There are two zones divided by dissimilar interatomic spacings $d_{1(2)}$, which respectively can be classified as two homogeneous atomic chains. An atom located at the interface of the dissimilar array is denoted as the interface atom with two disparate nearest-neighbor spacings. The inset plot shows the two-level quantum emitter with $\ket{g}$ and $\ket{e}$ indicating the ground and excited states, respectively.}\label{fig1}
\end{figure}

\section{Theoretical model}

We consider a weakly-driven chirally-coupled atomic array consisting of atoms that are uniformly spaced at $d_{1}$ and $d_{2}$ in the left and right segments, respectively, as illustrated in Fig. \ref{fig1}. The atom located at the intersection of the two segments is termed the ``interface atom", which bridges two segments with two different spacings $d_{1}\neq d_{2}$. Its role will be clear and crucial in the following discussions throughout the paper, since it can be treated as belonged to either side of the dissimilar array and would modify the characterization of the steady-state phases in the system in some parameter regimes. 

The effective model of an atom-waveguide system can be obtained by treating the guided modes on a waveguide as a one-dimensional reservoir \cite{Tudela2013, Pichler2015}. Within the interaction picture and under the Born-Markov approximation \cite{Lehmberg1970}, the dynamical evolution of the system's density matrix $\rho$ can be governed by 
\begin{equation}\label{chiral master eq}
    \dv{\rho}{t}
    =
    -\dfrac{i}{\hslash}[H_{\text{S}}+H_{\text{L}}+H_{\text{R}},\rho]
    +
    \mathcal{L}_{\text{L}}[\rho]
    +\mathcal{L}_{\text{R}}[\rho], 
\end{equation}
with Hamiltonians $H_{\text{S}}$ the light-matter interaction from a laser excitation, $H_{\rm L(R)}$ the chiral couplings, and Lindblad forms of $\mathcal{L}_{\text{L(R)}}[\rho]$ indicating the chiral dissipations to the left (L) or the right (R) side of the waveguide. The Hamiltonian $H_{\text{S}}$ is 
\begin{equation}
    H_{\text{S}}
    =
    \hslash
    \sum_{\mu=1}^{N}\left[\Omega e^{ikx_{\mu}\cos\theta }\left(
    \sigma_{\mu}+\sigma_{\mu}^\dagger\right)
    -\delta_{\mu}\sigma_{\mu}^\dagger\sigma_{\mu}\right], 
\end{equation}
which drives a collection of $N$ two-level quantum emitters (characterized by the ground state $\ket{g}$ and the excited state $\ket{e}$) with a Rabi frequency $\Omega$ and spatially dependent detunings $\delta_j$. The dipole raising and lowering operators are defined as $\sigma_\mu^\dagger \equiv \ket{e}_{\mu}\bra{g}$ and $\sigma_{\mu} = (\sigma_{\mu}^\dagger)^\dagger$, respectively. $k=2\pi/\lambda$ is the wave number with the transition wavelength $\lambda$, while the uniform excitation angle denoted as $\theta$ characterizes the propagation phases of the driving field from a lateral excitation away but close to the $\hat{x}$-$\hat{z}$ plane. The $H_{\text{L(R)}}$ are
\begin{align}
    H_{\text{L(R)}}
    =
    -i\hslash\dfrac{\gamma_{\text{L(R)}}}{2}
    \sum_{\mu<(>)\nu}^{N}\left(e^{ik|x_{\mu}-x_{\nu}|}\sigma_{\mu}^\dagger\sigma_{\nu}-\text{h.c.}\right),
\end{align}
which denote the collective energy shifts arising from the infinite-range photon-mediated dipole-dipole interaction \cite{Solano2017}. The remaining dissipative parts in Lindblad forms are
\bea
    \mathcal{L}_{\text{L(R)}}[\rho]
    =\gamma_{\text{L(R)}}
    \sum_{\mu,\nu=1}^{N}e^{\mp ik(x_{\mu}-x_{\nu})}
    \left[
    \sigma_{\nu}\rho\sigma_{\mu}^\dagger
    -\dfrac{1}{2}\{\sigma_{\mu}^\dagger\sigma_{\nu},\rho\}
    \right],\nonumber\\
\eea
which describe the collective and nonreciprocal decay rates with $\gamma_{\textrm L}\neq\gamma_{\textrm R}$ in general. The directionality factor of the couplings can be quantified as $D\equiv (\gamma_R - \gamma_L)/\gamma$ \cite{Mitsch2014} with $\gamma = \gamma_R + \gamma_L \equiv 2\vert dq(\omega)/d\omega\vert_{\omega = \omega_{eg}}g^2_{k}L$ \cite{Tudela2013}. $\vert dq(\omega)/d\omega\vert$ indicates the inverse of group velocity at a resonant wavevector $q(\omega)$, with the atom-waveguide coupling strength $g_{k_s}$ and the quantization length $L$. 

We note that the Lindblad forms presented above neglect the nonguided decay or additional nonradiative losses experienced by the atoms. This omission could potentially affect the efficiency of light detection through fibers, as well as the fidelity of the steady-state preparations. We also label the positions of atoms such that $x_{\mu}>x_{\nu}$ when $\mu>\nu$ for the array without loss of generality. To facilitate our analysis, we initialize the system in the ground state $\ket{g}^{\otimes N}$ and consider the system dynamics under a weak-field excitation. This results in a confined and self-consistent Hilbert subspace for dynamical evolutions within the ground state and singly-excited states $\{\ket{g}^{\otimes N}, \ket{\psi_{\mu}}\in\ket{e}_{\mu}\ket{g}^{\otimes (N-1)}\}$ for $\mu=[1,N]$. This leads to the state representation, 
\begin{equation}
\ket{\Psi(t)}=\dfrac{1}{\sqrt{1+\sum_{\mu=1}^N|p_{\mu}(t)|^2}}\left(\ket{g}^{\otimes N} +\sum_{\mu=1}^N p_{\mu}(t)\ket{\psi_{\mu}}\right),
\end{equation}
where $p_{\mu}(t)$ represents the probability amplitude and $\sum_{\mu=1}^N|p_{\mu}(t)|^2\ll 1$ to satisfy the assumption of a weak-field excitation. 

Thus Eq. (\ref{chiral master eq}) can be reduced to a non-Hermitian Schrödinger equation $i\hbar\partial_{t}\ket{\Psi(t)}=H_{\text{eff}}\ket{\Psi(t)}$ with the effective Hamiltonian $H_{\text{eff}}$, yielding the coupled equations for $p_{\mu}(t)$ as 
\begin{equation}\label{probability amplitude}
    \dot{p}_{\mu}=-i\Omega e^{ikx_{\mu}\cos\theta}+\sum_{\nu=1}^{N}M_{\mu\nu}p_{\nu}, 
\end{equation}
where the matrix elements $M_{\mu\nu}$ of the coupling matrix $M$ are 
\bea\label{coupling matrix}
M_{\mu,\nu} =
\begin{cases}
    -\gamma_L e^{ik|x_{\mu}-x_{\nu}|}&,\; \mu < \nu
    \\
    i\delta_{\mu}-\frac{\gamma_L+\gamma_R}{2}&,\; \mu = \nu
    \\
    -\gamma_R e^{ik|x_\mu-x_{\nu}|}&,\; \mu > \nu
\end{cases}. 
\eea
Consequently, the probability amplitudes in the steady states ($\dot{p}_{\mu}=0$) are given by
\begin{equation}\label{probability amplitude of steady state}
    \tilde{p}_{\mu}\equiv p_{\mu}(t\to\infty)
    =i\Omega \sum_{\nu=1}^{N}(M^{-1})_{\mu\nu}e^{ikx_{\nu}\cos\theta}.
\end{equation}
For convenience, we define the interatomic distances as 
\begin{equation}
k(x_{\mu}-x_{\mu-1})=
\begin{cases}
kd_{1}\equiv \xi_{1}&,\,1<\mu\leq m
\\
kd_{2}\equiv \xi_{2}&,\,m<\mu\leq N
\end{cases}
\end{equation}
where $m$ denotes the index of the interface atom. From Eqs. (\ref{probability amplitude}) and (\ref{coupling matrix}), we are able to identify the interaction-driven quantum phases of atomic excitations, which are predominantly determined by the directionality of couplings $D$ and photon-mediated dipole-dipole interactions. The latter are decisively influenced by the interatomic separations. In the following we proceed to characterize the composite quantum phases that emerge from a homogeneous atomic array and discuss the interaction-induced depletion that is unique and only exists in certain parameter regimes. 

\section{Steady-state phases}

Here we obtain the steady-state phases in a weakly-driven chirally-coupled dissimilar atomic array. Throughout the study, we locate the interface atom at the site of $m = \lceil N/2\rceil$ with a ceiling function. Since $D$ quantifies the degree of directional couplings and dissipations, we denote $D=0$ as bidirectional and $\pm1$ as unidirectional coupling, respectively. We assume a perpendicular laser excitation at $\theta = \pi/2$ and uniform laser detunings $\delta_{\mu} = 0$. Under these conditions, we numerically calculate the normalized steady-state population distributions $\tilde{P}_{j}$ which are defined as:
\begin{equation}
    \tilde{P}_{j}=\dfrac{\left|\tilde{p}_{j}\right|^2}{\sum_{j=1}^{N}\left|\tilde{p}_{j}\right|^2}.
\end{equation}
For the case of a homogeneous atomic array, we have identified the steady-state phase diagram under the parameters of $D$ and interparticle distance $\xi$ \cite{Jen2020_phase}. In this weakly-driven atomic array, the steady states can be characterized as the extended distributions (ETD) when $\xi\approx 0$, the phase with finite crystalline (CO) orders, the bi-edge excitations (BE), the bi-hole excitations (BH), and the chiral-flow dichotomy (CFD) when $\xi=\pi$. The ETD, CO, and BH phases present the delocalized characteristics, while the BE phase manifests localization properties, which can be distinguished by the participation ratio \cite{Murphy2011}. Meanwhile, the CO phase possesses an extra finite structure factor and the BH phase shows hole excitations at the edges, which further differentiates the delocalized phases apart. The CFD phase resides at a very specific parameter of $\xi=\pi$ and shows two different population distributions depending on an odd or even $N$. This presents another interesting steady-state phase of the excited-state populations, where an extra number of atom gives rise to contrasted phases, and it resonates with the notion of ``More is different" owing to the complexity and hierarchical structure in interacting quantum systems \cite{Anderson1972}. Two other critical parameter regimes of $D=0$ and $\xi = \{0,\pi\}$ can also be identified and excluded from the steady-state phases owing to the divergence in state populations, a feature of decoherence-free space and therefore leading to a breakdown in the weak-excitation assumption.

\begin{figure}[t]
\includegraphics[width=0.48\textwidth]{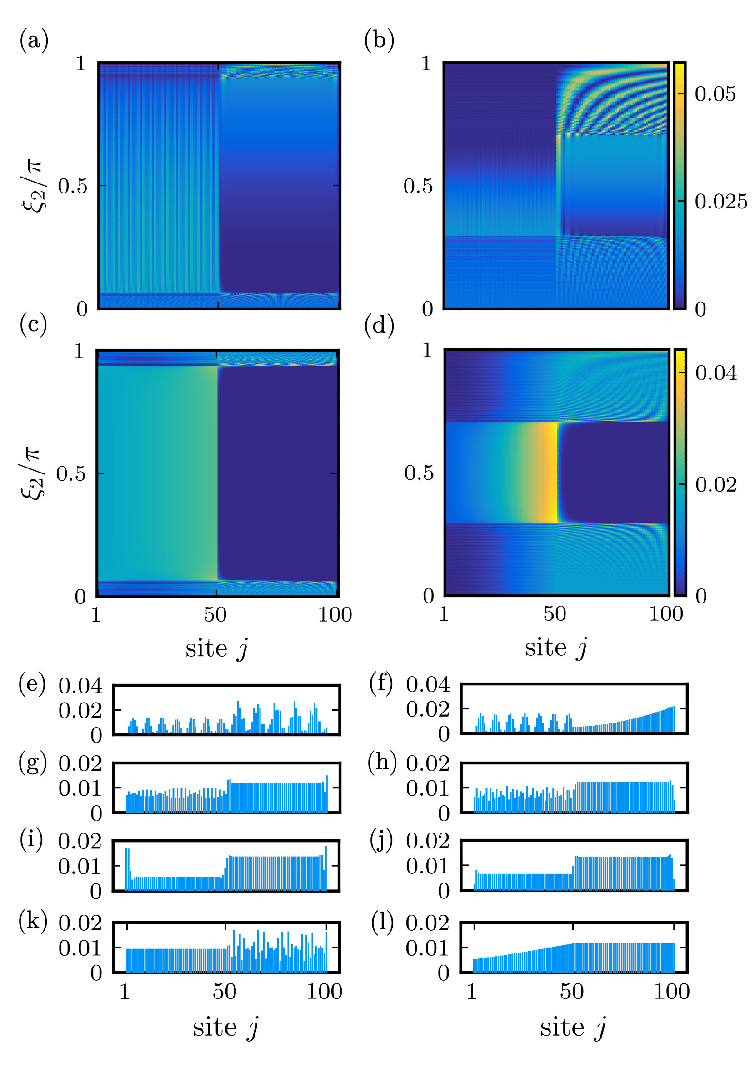}
\caption{The population distributions $\tilde P_j$ of the dissimilar array for $N=100$ atoms. We illustrate $\tilde P_j$ with a dependence of $\xi_2$ at some chosen $\xi_1$ and $D$. In the panels (a-d), we set $\xi_1=0.02\pi$ in (a) and (b), and $\xi_1=\pi$ in (c) and (d), while we choose $D=0.2$ in (a) and (c), and $D=0.8$ in (b) and (d), as comparisons. Some combinations of steady-state phases are demonstrated: (e) and (f) at $D=0.5$ display CO-CO phase ($\xi_1=0.15\pi, \xi_2=0.9\pi$) and CO-CFD phase ($\xi_1=0.15\pi, \xi_2=\pi$), respectively. (g) depicts CO-BE phase ($D=0.3, \xi_1=0.01\pi, \xi_2=0.3\pi$). (h) illustrates CO-BH phase ($D=0.5, \xi_1=0.1\pi, \xi_2=0.6\pi$). (i) and (j) at $D=0.4$ feature BE-HE phase ($\xi_1=0.2\pi, \xi_2=0.3\pi$) and BH-BH phase ($\xi_1=0.6\pi, \xi_2=0.7\pi$), respectively. (k) shows ETD-CO phase ($D=0.6, \xi_1=10^{-4}\pi, \xi_2=0.1\pi$). Finally, (l) presents CFD-ETD phase ($D=0.2, \xi_1=\pi, \xi_2=10^{-4}\pi$).}\label{fig2_population}
\end{figure}

In Fig. \ref{fig2_population}, we present the excited-state population distributions in terms of parameter spaces $\xi_1$, $\xi_2$, and $D$. Based on the results of steady-state phases in a homogeneous array, in the setup of a dissimilar array we have $17$ possible phases by accounting for the reflection symmetry in the system, which we discuss in detail in Appendix A. Interestingly, among these $17$ steady states, we identify two extra configurations termed Edge-Hole (EH) or Hole-Edge (HE) excitations, which results from the influence of the interface atom and distinguishes from the previous BE or BH excitation phases in a homogeneous array. Figures \ref{fig2_population}(a)-\ref{fig2_population}(d) show some examples of cross sections in the three-dimensional parameters of $\xi_1$, $\xi_2$, and $D$. At a smaller $\xi_1$ in Figs. \ref{fig2_population}(a) and \ref{fig2_population}(b), the left segment hosts the CO phase, while in the right segment of the array, the steady-state phases transition from the ETD phase, the CO phase, to the BE excitation, the BH excitation phases, then the CO phase, and end up with the CFD phase as $\xi_2$ increases from $0$ toward $\pi$. These six steady-state phases resemble parts of the features in a homogeneous array, and they can be seen clearly in either the left or the right segments of the array in Figs. \ref{fig2_population}(e)-\ref{fig2_population}(l). The dark-blue area in the right segment represents a low excited-state population in the bulk with the bi-edge excitations, while the light-blue area indicates a finite and flat distribution in the bulk, signifying the hole excitations at both the edges of the right segment. 

Similarly in Figs. \ref{fig2_population}(c) and \ref{fig2_population}(d), the left segment of the array hosts the CFD phase with corresponding phase transitions in the right segment as $\xi_2$ increases. As $D$ increases, we find the phase regions that host BE and BH excitations (dark and light blue regions in the right segment) are shrinking, which is expected and also appears in the case of homogeneous array \cite{Jen2020_phase}, along with expanded phase areas of the CO phases (interference patterns) in Figs. \ref{fig2_population}(b) and \ref{fig2_population}(d). We note that the population distribution can exhibit significant differences in the respective total populations of the two segments, as demonstrated in Fig. \ref{fig2_population}(a)-(d), with this disparity becoming particularly pronounced when $\xi_1=\pi$ in panels (c) and (d). This is why we see a seemingly flat distribution with low populations in the CO phase of the left segment in Fig. \ref{fig2_population}(b) (upper-left phase region). This manifests an intriguing steady-state phase with extremely low populations but still with the characteristics or the features of structure factors required in the CO phase. This low population can also be observed in the dark-blue areas in the right segment of Figs. \ref{fig2_population}(b) and \ref{fig2_population}(d), which we will study more in the next section. These extremely low population phase areas showcase a surprisingly and notably blockaded region of atomic excitations. A thorough investigation of the mechanisms driving this phenomenon and its impact on the population distribution will be investigated in the next section.

In Figs. \ref{fig2_population}(e)-\ref{fig2_population}(l), we show the excited-state populations for some combinations of the steady-state phases. For example in Fig. \ref{fig2_population}(e), two CO phases can be identified with finite and peak values of the structure factors at two different periods of excitation oscillations. Specifically, regarding Fig. \ref{fig2_population}(i), the expectation from the insights of a homogeneous array suggests a BE-BE phase under the corresponding parameters. However, the numerical simulations reveal that it appears as the BE-HE phase instead, regardless of whether the interface atom is associated to the left or the right segment of the array. By contrast, when the number of atoms is odd (for example of $N=101$ with the interface atom at the site of $m=51$) and under the same parameters in Fig. \ref{fig2_population}(i), classifying the interface atom on the left segment of the atomic chain results in the BE-HE phase, while on the right segment leads to the EH-HE phase. This outcome persists as the EH-HE phase even when the interface atom is excluded from both segments. If the interface atom is considered as the shared element of both segments, the overall system can be described again back to the BE-HE configuration. It is evident that the right segment consistently exhibits the HE phase regardless of the classifications of the interface atom. Given that the interface atom straddles the boundary of two different interatomic distances, its affiliation to either the left or the right segments becomes adjustable. Therefore, in the scenarios with an odd number of atoms, we refer to this population distribution as the “overall BE" phase.  

The CFD phase displayed in the left segment of Fig. \ref{fig2_population}(l) varies depending on the parity of the atom number in the left. When its atom number is even, the state in the CFD phase presents a linearly-increasing shape, as evidenced as well in Figs. \ref{fig2_population}(b) and \ref{fig2_population}(d) with a phase region when $\xi_2$ is around $\pi/2$. Conversely, with an odd number of atoms in the left under the same parameters, it would exhibit an upwardly concave curve instead as in the right segment in Fig. \ref{fig2_population}(f). This phenomenon is linked to the classification of the interface atom in which side of the atomic array. Intriguingly, even in cases with an even total number of atoms, an odd-numbered CFD can manifest, and vice versa. Therefore, when $\xi_{i}=\pi$ for one segment in the CFD phase and $\xi_{j\neq i}$ on the other segment is chosen in the CO phase regime or near the boundary between the CO and the BE phases, this parameter regime triggers a sensitive alternation between even- and odd-numbered CFD phases. This can be seen in the distributions of the CFD phase in the left segment with striped patterns in Figs. \ref{fig2_population}(b) and \ref{fig2_population}(d), which are strongly associated with the CO phases in the right. This presents an intricate steady-state behaviors of the system owing to the interplay between the number of atoms in each segments and the photon-mediated spin-exchange couplings determined by interparticle spacings.

\section{Interaction-induced depletion}

\begin{figure}[t]
    \centering
    \includegraphics[width=0.48\textwidth]{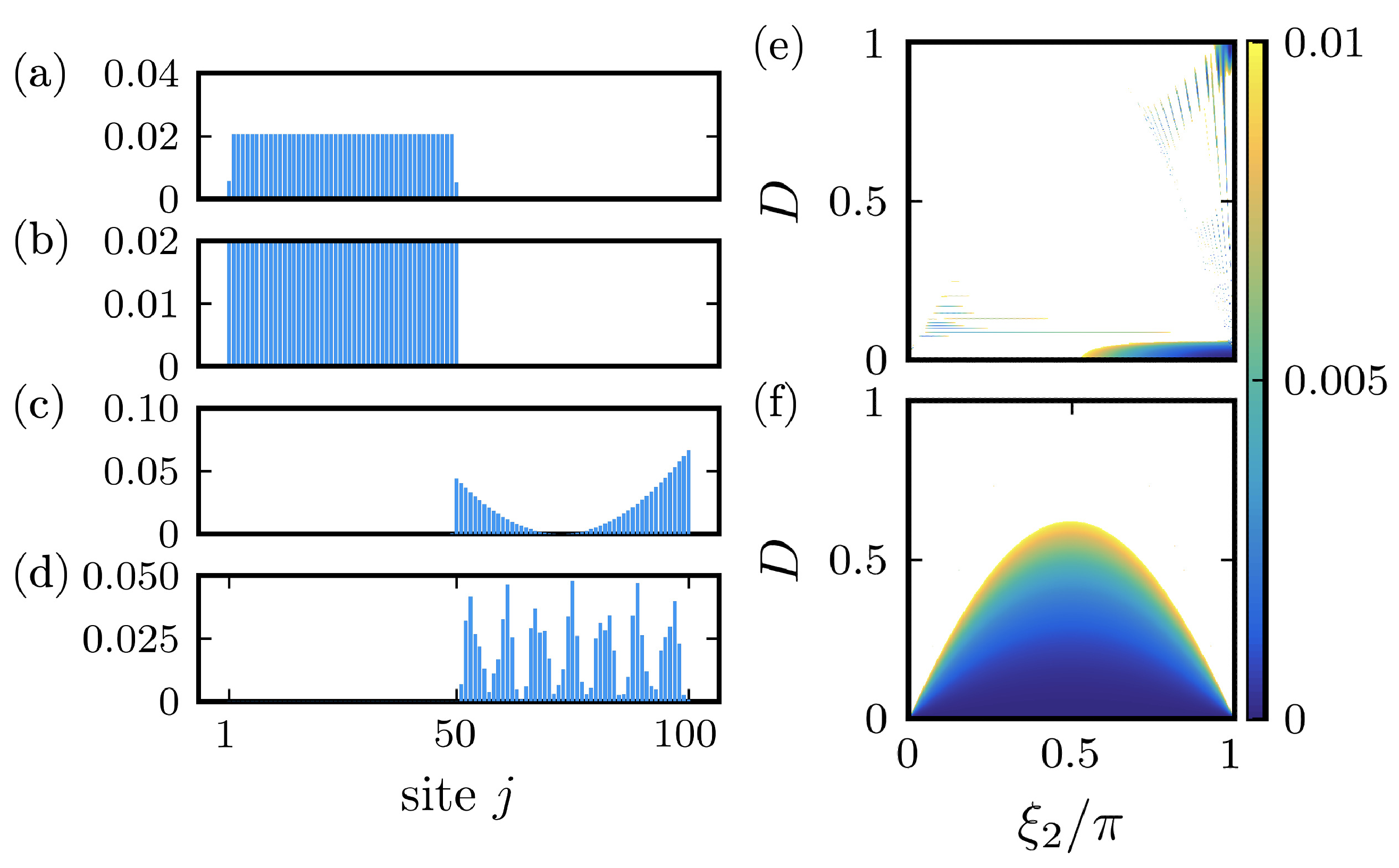}
    \caption{The interaction-induced half-depletion (HF) phases for $N=100$. (a) The BH-HD phase with mostly depleted ETD in the right segment under the parameters $D=0$, $\xi_1=0.8\pi$, and $\xi_2=0.02\pi$, with $B_D \approx 0.0013$. (b) The ETD-HD phase (depleted BE distribution in the right) for $D=0.1$, $\xi_1=10^{-5}\pi$, and $\xi_2=0.3\pi$, with $B_D \approx 0.0049$. (c) The HD-CFD phase (depleted Hole-Edge distributions in the left) for $D=0.2$, $\xi_1=0.5\pi$, and $\xi_2=\pi$, with $B_D \approx 0.0065$. (d) The HD-CO phase (depleted BE distribution in the left) for $D=0.4$, $\xi_1=0.2\pi$, and $\xi_2=0.9\pi$, with $B_D \approx 0.0018$. (e) and (f) represent the half-depletion regimes under $B_D \leq 1/N$, with fixed parameters of $\xi_1 = 0.02\pi$ and $\xi_1 = \pi$.}\label{figure3 depletion phases}
\end{figure}

As shown in Figs.\ref{fig2_population}(c) and \ref{fig2_population}(d), we observe a significant decline of the excitation population in one segment. We term this unexpected phenomenon as the depletion phase. The occurrence of the phase depletion can be attributed to the complex spin-exchange interactions among the atomic array, arising from variations in interparticle spacings of the dissimilar array. In Fig. \ref{figure3 depletion phases}(a)-\ref{figure3 depletion phases}(d), we illustrate varied half-depletion patterns in the system. These diverse depletion phenomena in half of the atomic array underline an intricate interplay with competing parameters of long-range dipole-dipole interactions and directionality of couplings. 

To systematically identify the parameter regimes that support this depletion phenomenon, we define a quantity of biased population ($B_D$) as a measure of the contrasted steady-state distributions between each segments of the dissimilar array, 
\begin{equation}\label{BD formula}
B_D \equiv 1-\left|\dfrac{\sum_{\mu=1}^{m-1}\tilde{P}_{\mu}-\sum_{\mu=m+1}^{N}\tilde{P}_{\mu}}{\sum_{\mu=1}^{N}\tilde{P}_{\mu}-\tilde{P}_m}\right|,
\end{equation}
where we exclude the contribution from the interface atom at the $m$th site. As our discussion primarily focuses on the extent of overall population bias towards one side of the atomic array, this exclusion does not modify the conclusion from the analysis. From Eq. (\ref{BD formula}), $B_D$ approaches one when a uniform distribution in the whole array is reached, while $B_D$ becomes vanishing as a depleted region emerges. We then characterize the half-depletion (HD) phase in a dissimilar array if and only if $B_D \leq 1/N$. That is to say, the total population of the depleted side must be lower than a single-particle population on average, which sets a qualitative boundary for the HD phase. 

In Figs. \ref{figure3 depletion phases}(e) and \ref{figure3 depletion phases}(f), we show the HD phase regime with two different $\xi_1$ for various $\xi_2$ and $D$ under this criteria. A dominant parameter regime for the HD phase appears at a lower $D$ for a small $\xi_{1}$ as shown in Fig. \ref{figure3 depletion phases}(e). As $\xi_1$ increases toward $\pi/2$, these parameter regions begin compressing towards a larger $\xi_2$ and extend along a higher $D$, which disappear as $\xi_1$ further increases along with new small regions of parameters emerging on the other side at small $\xi_2$. At $\xi_1$ reaches $\pi$, the HD phase boundary emerges with a parabola as seen in Fig. \ref{figure3 depletion phases}(f). This suggests that the depletion regime requires contrasted $\xi_1$ and $\xi_2$. 

\begin{figure}[t]
    \centering
    \includegraphics[width=0.43\textwidth]{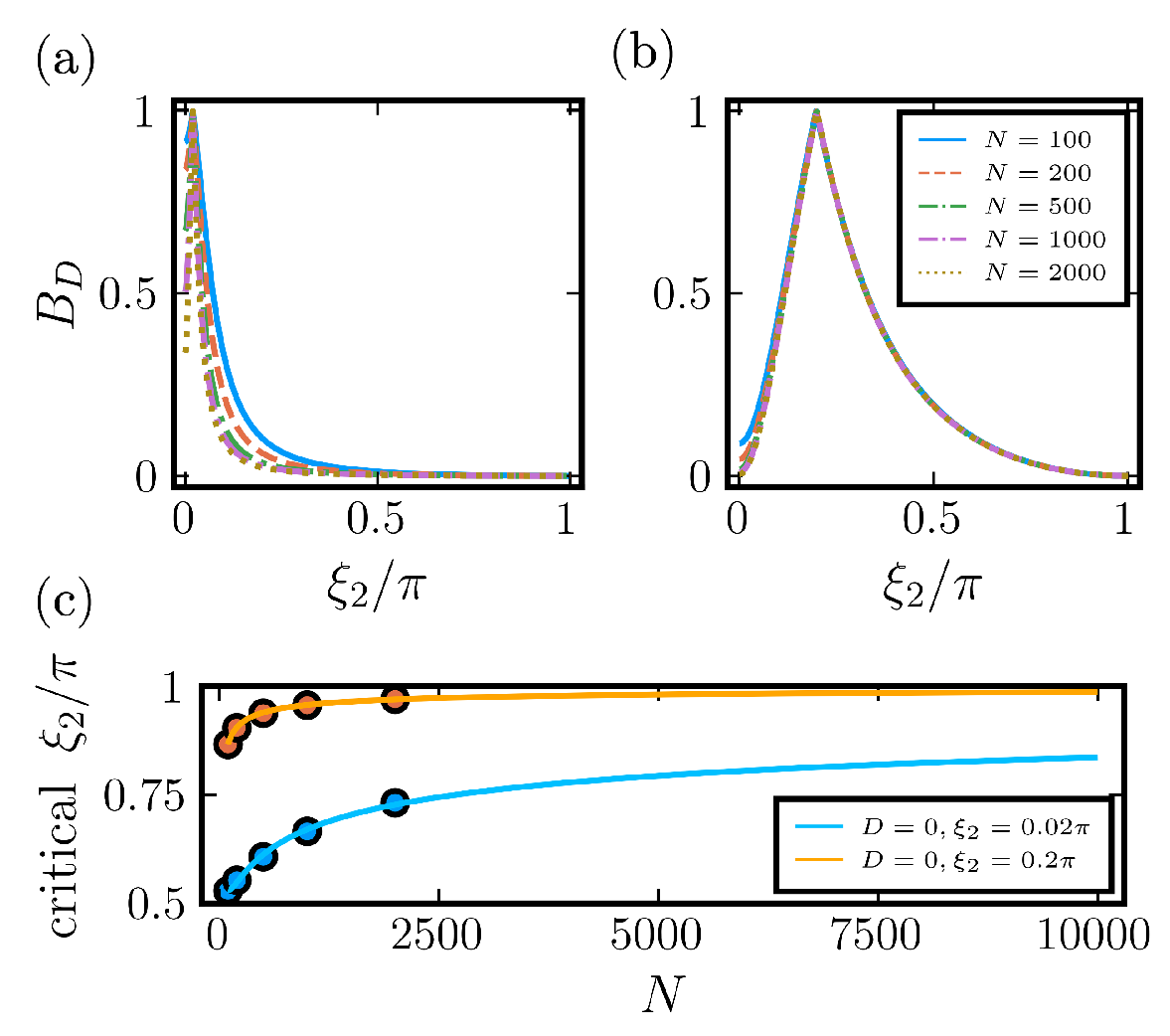}
    \caption{Impact of system size on the depletion regime in the bi-directional ($D=0$) case. The plots in (a) and (b) depict the dependence of $B_D$ on $\xi_2$ at $\xi_1=0.02\pi$ and $\xi_1=0.2\pi$, respectively, for various atomic sizes $N=100$, $200$, $500$, $1000$, and $2000$. (c) illustrates a fitted curve through five data points, each extracted as the critical $\xi_2$ to cross into the half-depletion regime for different $N$. Both scenarios of $\xi_2=0.02\pi$ (blue line) and $\xi_1=0.2\pi$ (orange line) reveal that the depletion phase is a consequence of finite-size effect, which do not survive as $N\rightarrow\infty$.}\label{figure4 BD_depletion}
\end{figure}

\begin{table}[b]
\centering
\begin{tabular}{cllcccc}
\\\hline\hline
\multicolumn{1}{l}{} &  &  & \multicolumn{4}{c}{critical $\xi_2/\pi$}     \\ \cline{4-7} 
$N$                  &  &  & at $\xi_1=0.02\pi$ &  &  & at $\xi_1=0.2\pi$ \\ \hline
100                  &  &  & 0.529              &  &  & 0.976             \\
200                  &  &  & 0.553              &  &  & 0.984             \\
500                  &  &  & 0.607              &  &  & 0.990             \\
1000                 &  &  & 0.665              &  &  & 0.992             \\
2000                 &  &  & 0.731              &  &  & 0.994             \\\hline\hline
\end{tabular}
\caption{Critical $\xi_2$ for $\xi_1=0.02\pi$ or $0.2\pi$.}
\label{critical xi2}
\end{table}

This can be evidenced in the bi-directional case with $D=0$ as shown in Figs. \ref{figure4 BD_depletion}(a) and \ref{figure4 BD_depletion}(b) with $\xi_1=0.02\pi$ and $0.2\pi$, respectively. The measure $B_D$ first rises as $\xi_2$ increases from $0$ to approximately $\xi_1$, reaching near $1$ when $\xi_1=\xi_2$, the case of a homogeneous array. As $\xi_2$ exceeds $\xi_1$ and increases further, $B_D$ rapidly decreases and becomes vanishing toward $\xi_2 = \pi$. Additionally, as illustrated in Fig. \ref{figure4 BD_depletion}(a), $B_D$ decreases faster with increasing $N$, especially in the case of a small $\xi_1$. This leads to a diminishing depletion region as $N$ increases, since the critical values of $\xi_2$ at which $B_D=1/N$ increase, indicating a shrinking phase area for the HD phase. We further analyze these values by fitting the critical $\xi_2$ in Table \ref{critical xi2}, beyond which the system enters the HD phase. In Fig. \ref{figure4 BD_depletion}(c), we utilize a fitting function of $(aN^b+cN^d+1)$ for critical $\xi_2/\pi$, where we obtain $(a,b,c,d)=(-29.4, -0.46, 35.6, -0.53)$ and $(-1.13, -0.49, -0.14, -0.49)$ for Figs. \ref{figure4 BD_depletion}(a) and \ref{figure4 BD_depletion}(b), respectively. The fitted asymptotic function indicates that the phase boundary $\xi_2\to\pi$ as $N\to\infty$, which manifests an ultimately vanishing depletion region. This also suggests that the depletion phenomenon under our restricted criteria results from the finite-size effect, except for the case of $\xi_{1}=\pi$ where all $\xi_2$ with a finite region of $D$ allow the existence of HD phase. As the criteria relaxes to the two-particle population on average, for example, or more but finite, the phase areas of the HD phase would expand and the fate of this phase in the thermodynamics limit could survive. Below we provide analytical solutions in the reciprocal coupling case of $D=0$, which can give insights to the mechanism of the HD phase.   

\subsection{\label{AnalyticalExplanation}Analytical solution in the bi-directional case}

\begin{figure}[b]
  \includegraphics[width=0.47\textwidth]{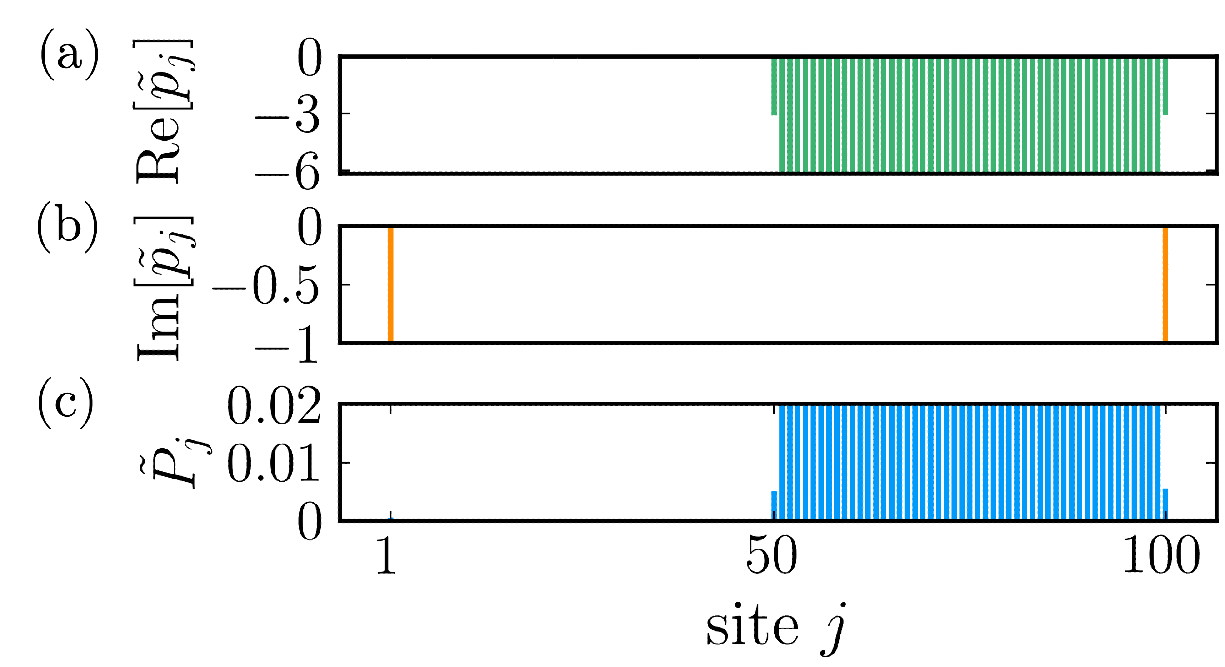}
	\caption{The imaginary (a) and real (b) parts of the probability amplitudes $\tilde{p}_{j}$ for $N=100$, $D=0$, $\xi_1=0.02\pi$, and $\xi_2 = 0.8\pi$. The values are up to an overall constant of $2\Omega$. The panel (c) shows the normalized excited-state population $\tilde P_j$.} 
  \label{Sigma&P}
\end{figure}

Under the bi-directional case of $D=0$, that is $\gamma_{\textrm R}=\gamma_{\textrm L}$, we obtain the steady-state populations $\tilde{p}_\mu$ from Eq. (\ref{probability amplitude of steady state}) as 
\begin{equation}
\tilde{p}_{\mu} = -\Omega\times
\begin{cases}
    i+\tan(\xi_1/2)
    &,\;\mu = 1
    \\
    2\tan(\xi_1/2)
    &,\;1 < \mu < m
    \\
    \tan(\xi_1/2)+\tan(\xi_2/2)
    &,\;\mu = m
    \\
    2\tan(\xi_2/2)
    &,\;m < \mu < N
    \\
    i+\tan(\xi_2/2)
    &,\;\mu = N
\end{cases},\label{bi}
\end{equation}
where the derivation is shown in detail in the Appendix B. A numerical calculation of $\tilde{p}_{j}$ for the parameters of $N = 100$, $D = 0$, $\xi_1 = 0.02\pi$, and $\xi_2 = 0.8\pi$ is plotted in Figs. \ref{Sigma&P}(a) and \ref{Sigma&P}(b), where the system hosts the HD phase as shown in Fig. \ref{Sigma&P}(c). These results are consistent with the theoretical predictions in Eq. (\ref{bi}) in both real and imaginary parts. 

From the measure $B_D$, we can analytically determine at which parameters the HD phase boundary locates. Considering an even $N$ with the interface atom at the site $m=N/2$, as in the numerical simulations shown in Fig. \ref{figure4 BD_depletion}(c) and for a small $\xi_1$ as in Fig. \ref{figure3 depletion phases}(e), we obtain the range of $\xi_2$ when the HD phase emerges from Eq. (\ref{BD2}),  
\begin{equation}
    \tan^2\left(\dfrac{\xi_2}{2}\right)\geq\frac{(N^2-2N+\frac{7}{4})\tan^2(\xi_1/2)+\frac{N-1}{2}}{\frac{N}{2}-\frac{3}{4}}.
\end{equation}
The phenomenon of depletion in the bi-directional cases straightforwardly occurs when $\tan^2(\xi_1/2)/\tan^2(\xi_2/2)\gtrsim 2N\ll 1$ (depleted from left side) or $\gg 1$ (depleted from right side) for a large $N$. This relation indicates a drastic difference between $\xi_1$ and $\xi_2$, which allows the HD phase under the criteria of $B_D\leq 1/N$. This coincides with a relatively large critical $\xi_2$ at the HD phase boundary in Figs. \ref{figure3 depletion phases} and \ref{figure4 BD_depletion}, respectively, for a low $D$.   

\section{Discussion and conclusion}

The study of atomic dissimilar array coupled to a nanophotonic waveguide under a weakly-driven condition provides one of the unique driven-dissipative quantum systems that can host many fascinating nonequilibrium dynamics and steady-state phases. This is due to an intricate interplay between various competing parameters of long-range dipole-dipole interactions and the directionality of couplings. Our investigation of such system reveals the essential combinations of steady-state phases from a homogeneous array, where new state configurations like a hole with an edge excitation at the interface and the edge, respectively, can emerge. In addition, the interface atom that bridges two segments with different interparticle spacings presents an intriguing role in deciding and characterizing the steady-state phases. Another intriguing and significant result is the apparent decline in state populations in one of the two segments, indicating a blockaded region of atomic excitations. This effect arises from the contrasted interparticle spacings in the dissimilar array near the reciprocal coupling regime, which we attribute as the interaction-induced half-depletion phase. Our results can provide insights to quantum engineering or quantum simulations of exotic many-body states with high controllability. 

The decrease in population distribution inside one of the segments has significant implications for quantum information processing. The depletion phase predicted in a specific segment adds to the complexity of the quantum state by introducing more degrees of freedom for encoding quantum information. This can be done in designing distinct arrangement of atoms, particularly when separated into two regions. The disparity in the interparticle distances between atoms then serves as a means to encode and manipulate quantum information in these states. For example, a single photon can go through a beam splitter and interact with two settings of dissimilar atomic arrays with either the left or the right segment hosts a depleted region. This leads to an entangled state of $|0_L1_R\rangle+|1_L0_R\rangle$ in terms of vanishing or occupied qubit states by constructing the entangling degrees of freedom in segmented spaces. The application of manipulating dissimilar atomic array thus promises many other opportunities in routing photons \cite{Yudi2023}, allowing parallel quantum operations \cite{Bekenstein2020}, or creating multipartite entangled states \cite{Chien2023}. 

\begin{figure*}[t]
  \includegraphics[width=0.9\textwidth]{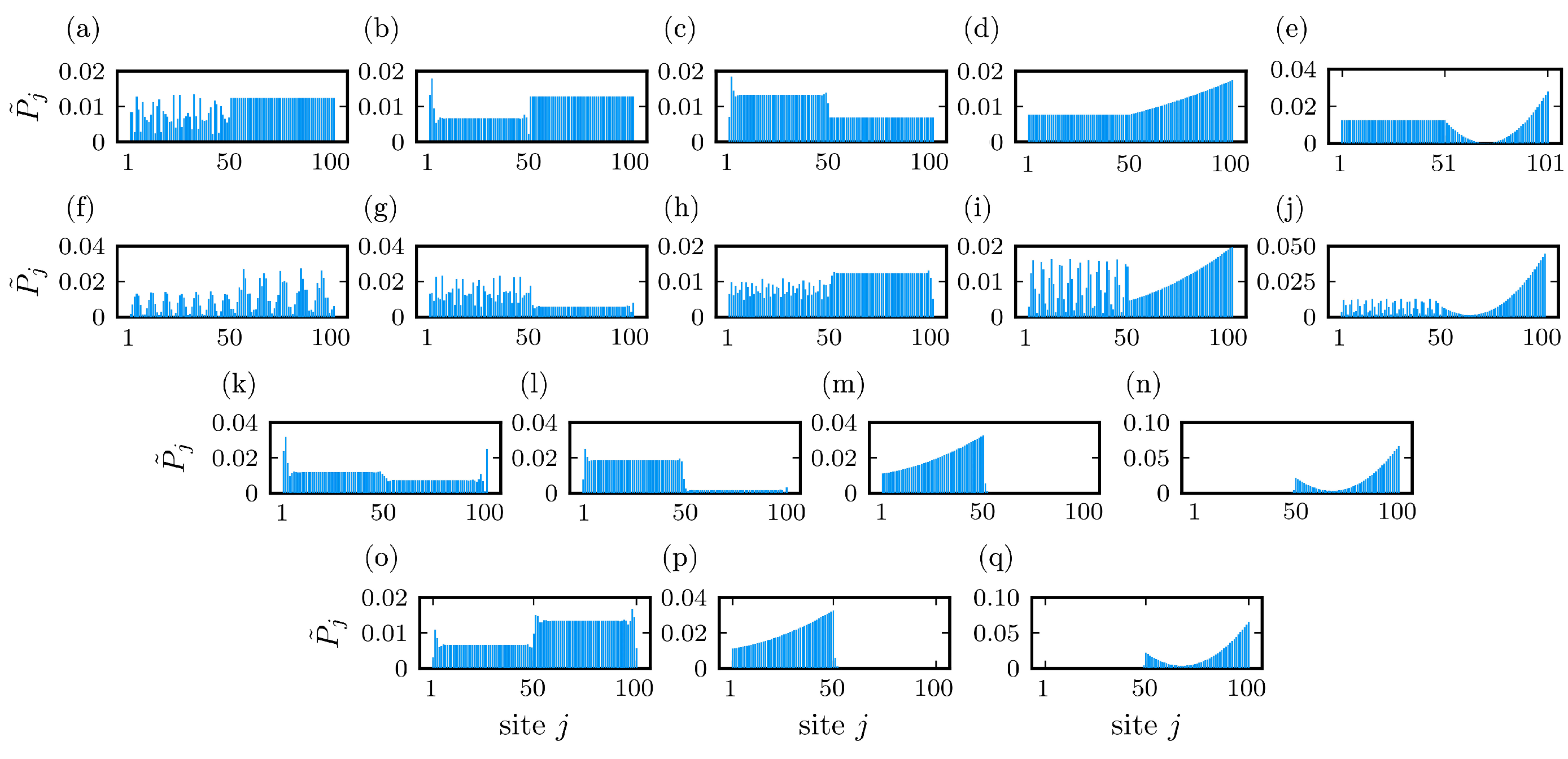}
	\caption{Excited-state distributions $\tilde P_j$ for seventeen steady-state phases in a dissimilar array. These phases include (a) CO-ETD, (b) BE-ETD, (c) BH-ETD, (d) ETD with even-numbered CFD, (e) ETD with odd-numbered CFD specifically at $N=101$, (f) CO-CO, (g) CO-BE, (h) CO-BH, (i) CO with even-numbered CFD, (j) CO with odd-numbered CFD, (k) EH-HE (overall BE), (l) BH-BE, (m) even-numbered CFD with BE, (n) BE with odd-numbered CFD, (o) BH-BH, (p) even-numbered CFD with EH, (q) HE with odd-numbered CFD. All phases are presented at $N=100$ except for (e) at $N=101$. The parameters for each panel are detailed in Table \ref{parameter_phases_state}.}
\label{fig6_17phases} 
\end{figure*}
\section*{ACKNOWLEDGMENTS}
We acknowledge support from the National Science and Technology Council (NSTC), Taiwan, under Grants No. 112-2112-M-001-079-MY3 and No. NSTC-112-2119-M-001-007, and from Academia Sinica under Grant AS-CDA-113-M04. We are also grateful for support from TG 1.2 of NCTS at NTU.
\appendix
\section{A complete set of steady-state phases in a dissimilar array}

Here we show the probability distributions for the complete seventeen steady-state phases in a dissimilar array. As shown in Fig. \ref{fig6_17phases}, we observe various combinations of steady-state phases in a homogeneous array. Other than these combinations, we have extra phases that a homogeneous array does not host, which are EH and HE configurations. Furthermore, we obtain the half-depletion phases that show extremely low state populations as shown in Figs. \ref{fig6_17phases}(l), \ref{fig6_17phases}(m), \ref{fig6_17phases}(n), \ref{fig6_17phases}(p), and \ref{fig6_17phases}(q). 

We note that there is no ETD-ETD phase since it becomes the ETD phase exactly as in a homogeneous array. There are also no allowed combinations with either even-numbered CFD (eCFD) or odd-numbered CFD (oCFD) phases, since they hugely depend on the boundary effect from the parity of the atom numbers. By contrast, the CO-CO or BH-BH phases can emerge owing to the disparity of respective state populations of two segments, which distinguishes the case in a homogeneous array. 

\begin{table}[b]
\centering
\caption{Parameter $D$, $\xi_1$, and $\xi_2$ for 17 phases}
\begin{tabular}{cllcllcll l ll l}
\hline\hline
Figure \ref{fig6_17phases} &&&  Steady-state phases  &&& $D$ &&& $\xi_1  /\pi$ &&& $\xi_2 /\pi$ \\ \hline
(a) &&&  CO-ETD              &&& 0.5      &&& 0.1      &&& $10^{-4}$      \\
(b) &&&  BE-ETD              &&& 0.5      &&& 0.25      &&& $10^{-4}$      \\
(c) &&&  BH-ETD              &&& 0.5      &&& 0.5      &&& $10^{-4}$      \\
(d) &&&  ETD-eCFD  &&& 0.2      &&& $10^{-4}$      &&& 1      \\
(e) &&&  ETD-oCFD  &&& 0.2      &&& $10^{-4}$      &&& 1      \\
(f) &&&  CO-CO              &&& 0.5      &&& 0.15      &&& 0.9      \\
(g) &&&  CO-BE              &&& 0.5      &&& 0.1      &&& 0.3      \\
(h) &&&  CO-BH              &&& 0.5      &&& 0.1      &&& 0.6      \\
(i) &&&  CO-eCFD  &&& 0.4      &&& 0.1      &&& 1      \\
(j) &&&  CO-oCFD  &&& 0.5      &&& 0.1      &&& 1      \\
(k) &&& EH-HE              &&& 0.5      &&& 0.25      &&& 0.2      \\
(l) &&&  BH-BE              &&& 0.5      &&& 0.6      &&& 0.25      \\
(m) &&&  eCFD-BE  &&& 0.5      &&& 1      &&& 0.25      \\
(n) &&&  BE-oCFD  &&& 0.5      &&& 0.25      &&& 1      \\
(o) &&&  BH-BH              &&& 0.7      &&& 0.6      &&& 0.7      \\
(p) &&&  eCFD-EH  &&& 0.5      &&& 1      &&& 0.75      \\
(q) &&&  HE-oCFD  &&& 0.5      &&& 0.75      &&& 1      \\
\hline\hline
\end{tabular}
\label{parameter_phases_state}
\end{table}

\section{\label{Append.Derive}Derivation of excited-state probability amplitudes}

Here we provide some details of the derivation of the excited-state probability amplitudes. We first consider the homogeneous array with identical interparticle spacings. We consider the laser incident angle $\theta = \pi/2$ and detuning $\delta_{\mu}=0$, and we obtain the coupling matrix as (normalized to $\gamma$),
\begin{equation}
M_{\mu\nu} =-\frac{1}{2}\times 
\begin{cases}
    (1-D) e^{i\xi|\nu-\mu|}&,\;\mu < \nu\\
    1 &,\;\mu = \nu\\
    (1+D) e^{i\xi|\mu-\nu|}&,\;\mu > \nu\\
\end{cases}.
\end{equation}
The inverse of $M$ in the bi-directional coupling case can then be obtained as
\begin{equation}
\begin{split}
    &(M^{-1})_{\mu\nu} =
    \frac{-2}{1-e^{2i\xi}}\times
    \begin{cases}
        1            
        &,\;\mu = \nu = 1 \text{ or } N
        \\
        1+e^{2i\xi}   
        &,\;1 < \mu = \nu < N 
        \\
        -e^{i\xi}     
        &,\;|\mu-\nu| = 1
        \\
        0             
        &,\;\text{otherwise}
    \end{cases}
    ,
\end{split}
\end{equation}
and from Eq. (\ref{probability amplitude}), we obtain the steady-state solutions of the dissimilar array, 
\begin{equation}
    \tilde{p}_{\mu}
    =-\Omega\times
    \begin{cases}
        i+\tan(\xi_1/2)
        &,\;\mu = 1 \text{ or } N
        \\
        2\tan(\xi/2)&,\;1 < \mu < N.
    \end{cases}
\end{equation}

Similarly, for the case of a dissimilar array, the coupling matrix $M$ when $D=0$ can be given as
\begin{equation}
    M_{\mu\nu} =-\frac{1}{2}\times
    \begin{cases}
        e^{i\xi_1|\nu-\mu|}&,\; \mu < \nu \leq m\\
        e^{i\xi_1|m-\mu|+i\xi_2|\nu-m|}&,\; \mu < m < \nu\\
        e^{i\xi_2|\nu-\mu|}&,\; m \leq \mu < \nu\\
        1&,\; \mu = \nu\\
        e^{i\xi_1|\nu-\mu|}&,\; m \geq \mu > \nu\\
        e^{i\xi_1|m-\nu|+i\xi_2|\mu-m|}&,\; \mu > m > \nu\\
        e^{i\xi_2|\mu-\nu|}& ,\;\mu > \nu \geq m\\
    \end{cases}.
\end{equation}
The inverse matrix $M^{-1}$ can therefore be written as
\begin{widetext}
    \begin{equation}
    M^{-1}=-2\times\begin{bmatrix}  \frac{1}{1-e^{2i\xi_1}} & \frac{-e^{i\xi_1}}{1-e^{2\xi_1}} & 0 & \cdots & \cdots & \cdots & \cdots & \cdots & 0\\  \frac{-e^{i\xi_1}}{1-e^{2\xi_1}}& \frac{1+e^{2i\xi_1}}{1-e^{2i\xi_1}} & \frac{-e^{i\xi_1}}{1-e^{2\xi_1}} & 0 & \cdots & \cdots & \cdots & \cdots & 0\\  0      & \frac{-e^{i\xi_1}}{1-e^{2\xi_1}} & \ddots & \ddots & 0 & \cdots & \cdots & \cdots & 0\\  \vdots & 0      & \ddots & \frac{1+e^{2i\xi_1}}{1-e^{2i\xi_1}} & \frac{-e^{i\xi_1}}{1-e^{2\xi_1}} & 0 & \cdots & \cdots & 0\\  \vdots & \vdots & 0      & \frac{-e^{i\xi_1}}{1-e^{2\xi_1}} & \frac{1-e^{2i(\xi_1+\xi_2)}}{(1-e^{2i\xi_1})(1-e^{2i\xi_2})} & \frac{-e^{i\xi_2}}{1-e^{2\xi_2}} & 0 & \cdots & 0\\  \vdots & \vdots & \vdots & 0      & \frac{-e^{i\xi_2}}{1-e^{2\xi_2}} & \frac{1+e^{2i\xi_2}}{1-e^{2i\xi_2}} & \frac{-e^{i\xi_2}}{1-e^{2\xi_2}} & \ddots & \vdots\\  \vdots & \vdots & \vdots & \vdots & 0      & \frac{-e^{i\xi_2}}{1-e^{2\xi_2}} & \ddots & \ddots & 0\\  \vdots & \vdots & \vdots & \vdots & \vdots & \ddots & \ddots & \frac{1+e^{2i\xi_2}}{1-e^{2i\xi_2}} & \frac{-e^{i\xi_2}}{1-e^{2\xi_2}} \\  0      & 0      & 0      & 0      & 0      & 0      & 0      & \frac{-e^{i\xi_2}}{1-e^{2\xi_2}}    & \frac{1}{1-e^{2i\xi_2}}\end{bmatrix}.
\end{equation}
\end{widetext}

By solving the inverse matrix of $M$, we have the probability amplitudes $\tilde{p}_{\mu}$,
\begin{equation}
\tilde{p}_{\mu} = -\Omega\times
\begin{cases}
    i+\tan(\xi_1/2)
    &,\;\mu = 1
    \\
    2\tan(\xi_1/2)
    &,\;1 < \mu < m
    \\
    \tan(\xi_1/2)+\tan(\xi_2/2)
    &,\;\mu = m
    \\
    2\tan(\xi_2/2)
    &,\;m < \mu < N
    \\
    i+\tan(\xi_2/2)
    &,\;\mu = N
\end{cases}
\end{equation}
where $\xi_{1,2}=0$ and $\pi$ should be excluded. Therefore, we obtain the biased population $B_D$ as 
\begin{equation}
    1
    -\left|\frac{\left(m-\frac{7}{4}\right)\tan^2\left(\frac{\xi_1}{2}\right)-\left(N-m-\frac{3}{4}\right)\tan^2\left(\frac{\xi_2}{2}\right)}{\frac{1}{2}+\left(m-\frac{7}{4}\right)\tan^2\left(\frac{\xi_1}{2}\right)+\left(N-m-\frac{3}{4}\right)\tan^2\left(\frac{\xi_2}{2}\right)}\right|.
\end{equation}
With an even $N$ and $m=N/2$, the $B_D$ becomes
\begin{equation}
B_D=1-\left|\frac{(\frac{N}{2}-\frac{7}{4})\tan^2\left(\frac{\xi_1}{2}\right)-(\frac{N}{2}-\frac{3}{4})\tan^2\left(\frac{\xi_2}{2}\right)}{\frac{1}{2}+(\frac{N}{2}-\frac{7}{4})\tan^2\left(\frac{\xi_1}{2}\right)+(\frac{N}{2}-\frac{3}{4})\tan^2\left(\frac{\xi_2}{2}\right)}\right|.\label{BD2}
\end{equation}

\end{document}